\documentclass[twocolumn,pre,showpacs]{revtex4}

\usepackage{graphicx}
\usepackage{rotating}
\usepackage{amsmath}
\usepackage{amsfonts}
\usepackage{amssymb}
\usepackage{enumerate}
\usepackage{longtable}
\usepackage{dcolumn}
\usepackage{bm}
\usepackage{epsfig}
\usepackage{morefloats}
\usepackage{xcolor}
\usepackage[utf8]{inputenc}

\usepackage{etex}
\reserveinserts{23}

\definecolor{change}{rgb}{0,0,1} 
\definecolor{comment}{rgb}{1,0,0}
\definecolor{question}{rgb}{0,1,0}

\begin{document}

\title{
Modelling community formation driven by the status of individual in a society}

\author{Jan E. Snellman$^{1}$}
\author{Gerardo I\~niguez$^{2,1,3}$}
\author{Tzipe Govezensky$^{4}$}
\author{R. A. Barrio$^{5}$}
\author{Kimmo K. Kaski$^{1}$}

\affiliation{
$^1$Department of Computer Science, Aalto University School of Science, FI-00076 AALTO, Finland,\email{jan.snellman@aalto.fi}}
\affiliation{
$^2$Instituto de Investigaciones en Matem\'aticas Aplicadas y en Sistemas, Universidad Nacional Aut\'onoma de M\'exico, 01000 M\'exico D.F., Mexico}
\affiliation{
$^3$Centro de Investigaci\'on y Docencia Econ\'omicas, Consejo Nacional de Ciencia y Tecnolog\'ia, 01210 M\'exico D.F., Mexico}
\affiliation{
$^4$Instituto de Investigaciones Biomédicas, Universidad Nacional Autónoma de México, 04510 México D.F., Mexico}
\affiliation{
$^5$Instituto de F\'isica, Universidad Nacional Aut\'onoma de M\'exico, 01000 M\'exico D.F., Mexico}

\date{\textrm{\today}}

\begin{abstract}
{In human societies, people's willingness to compete and strive for better social status as well as being envious of those perceived in some way superior lead to social structures that are intrinsically hierarchical. Here we propose an agent-based, network model to mimic the ranking behaviour of individuals and its possible repercussions in human society. The main ingredient of the model is the assumption that the relevant feature of social interactions is each individual's keenness to maximise 
his or her status relative to others. The social networks produced by the model are homophilous and assortative, as frequently observed in human communities and most of the network properties seem quite independent of its size.
However, it is seen that for small number of agents the resulting network consists of disjoint weakly connected communities while being highly assortative and homophilic. On the other hand larger networks turn out to be more cohesive with larger communities but less homophilic. We find that the reason for these changes is that larger network size allows agents to use new strategies for maximizing their social status  allowing for more diverse links between them. 
}
{community formation, opinion formation, social hierarchy}
\end{abstract}

\maketitle


\section{Introduction}
\label{intro}

One of the most pervasive tendencies of humans is putting things in ranking order. In human societies these tendencies are reflected in their social interactions and networks being hierarchical in many respects. Hierarchies and ranks emerge due to individuals' subjective perceptions that some other individuals are in some respect better. Then a relevant research question is whether or not the formation and structure of hierarchies in human societies can be understood by making the assumption that the dominant driving force of people in social interactions is to enhance their own ``value'' or ``status'' relative to others. We call this assumption ``better than-hypothesis'' (BTH) and note that it is closely related to the thinking of the school of individual psychology founded by Adler in the early 1900s \cite{adler}, which, while starting with the assumption that human individuals universally strive for ``superiority'' over others, emphasizes inferiority avoidance as a motive for many human actions. 

Further studies of this kind of individuals' status-seeking behaviour, especially concerning consumer behaviour and economics, include the canonical references by Veblen \cite{V1899}, Duesenberry \cite{D1949} and Packard \cite{P1959} 
(See also Refs \cite{EGF1999,WF1998,R2008,GS2008}). 
In addition there is a closely related sociological model called Social Dominance Theory \cite{SDT}, which proposes that the construction and preservation of social hierarchies is one of the main motivations of humans in their social interactions and networks. However, the most relevant observational facts concerning BTH come from the field of experimental economics, especially from the results of experiments on the so-called ``ultimatum game'' \cite{GSS1982}, where the human players have been shown to reject too unequal distributions of money. The concept of {\em inequity aversion}, that is the observed social phenomenon of humans preferring equal treatment in their societies, is often invoked to explain these observations. Recently some models featuring inequity aversion have been proposed in 
Refs. \cite{ES1999,BO2000}.

All of these models, although from different fields of study, have something to do with the relative standings between different human individuals and groups, and so they could all be considered to emerge from or be based on a single principle such as BTH. It is this generality which makes BTH an intriguing and interesting object of study. There are even some studies on economic data, such as \cite{L2005}, that suggest a link between relative social standings and human well-being, and considerations of social status have measurable effects on brain functions, as shown in e.g. \cite{ISS2008,ZTCB2008}. 
These studies imply that BTH could well be something fundamental to human nature.  

The competition for a better hierarchical position among humans can be intense and sometimes even violent. However, humans have other characteristics including egalitarianism as well as striving for fairness. These traits could be interpreted in the context of BTH by remarking that people need to live in societies and make diverse social bonds, which in turn would contribute to their social status. This means that the members of society when they make decisions, need to take the feelings of others into account. Hence the behavioral patterns of individuals in social networks should then be characterised by sensitivity to the status of the other individuals in the network. This sensitivity manifests itself as inequity aversion and treating others fairly. To find out what in this context are the plausible and relevant mechanisms of human sociality driving societal level community formation we will focus on improving the BTH-based approach by using the frame of agent-based models and studying the emergence of social norms in such social systems, following the tradition presented in Refs. \cite{A1981,A1986,Y1993,Y1995,CWM2005,FOCN2016}. 

In this study we use an agent-based network model applying BTH-based approach to simulate social interactions dependent on societal values and rank, to get insight to their global effects on the structure of society. We find that in such a model society with a given constant ranking system the social network forms a degree hierarchy on top of the ranking system under BTH, such that the agents' degrees tend to increase, the further away their rank is from the average. The structure of the paper is as follows. In Section \ref{Model} we motivate the basics of BTH using the simple and well-researched ultimatum game as an example, and in Section \ref{modelv1} we show how the findings from this can be utilised as a part of agent-based models. In Section \ref{nc} we present the numerical results of the simulations from the model, and in Section \ref{meanfield} we analyse them. The two final Sections discuss the possible interpretations of the results and present the conclusions.

\section{Better than-hypothesis and ultimatum game}
\label{Model}

In this section we describe the theoretical basis for our model. We start by analysing the ultimatum game first proposed in \cite{GSS1982}, as it allows us to derive a basic form for the social gain function in our model. The ultimatum game is a game with two players, where one player has the task to make a proposal to the other player about how a given sum of money should be divided between them. The second player then gets to choose if the proposal is acceptable or not; if it is, the money is divided as proposed. If not, neither player gets anything. Experiments show that humans playing this game normally do not accept deals that are perceived to be unfair, i.e. in situations in which the proposer gets too large a share of the money (see, e.g. Refs. \cite{T1988,GT1990,kysymys,metalyysi,FHSS1994,BZ1995}). 
This is a classic problem in the mainstream economics, where humans are assumed to be rational and, therefore, accept something rather than nothing.

We implement BTH in the ultimatum game by interpreting the money used in a deal as a way of comparing the status between the one who accepts the proposal (called from now on the accepter) and its proposer. We denote the change of ``status'' of the accepter as $\Delta_a$, which takes into account it's own monetary gain, and the gain in relation to the proposer. Therefore, the simplest expression for $\Delta_a$  is,
\begin{equation}
\begin{aligned}
\Delta_a = &R_a(t_1) - R_a(t_0) + [R_a(t_1) - R_p(t_1)]\\
&-[R_a(t_0) - R_p(t_0)] ,
\end{aligned}
\label{perusta}
\end{equation}
where $R_a(t)$ and $R_p(t)$ stand for  
the monetary reserves (in the context of the game) of the accepter and proposer, respectively, at time $t$, with $t_0$ being the time before the deal and $t_1$ the time after the deal. In terms of economic theory, $\Delta_a$ would be called the accepter's change of utility, which is ordinarily assumed to consist of the term $R_a(t_1) - R_a(t_0)$ or the absolute payoff of the accepter. The additional terms $[R_a(t_1) - R_p(t_1)]-[R_a(t_0) - R_p(t_0)]$ that stem from the BTH measuring the change in relative standings of the accepter and proposer. The actual BTH utility function for the accepter in the ultimatum game takes the form
\begin{equation}
U^{BTH}_{a} = R_a + [R_a - R_p].
\end{equation}

According to Eq. (\ref{perusta}), the accepter will refuse the deal for $\Delta_a < 0$, and will accept it for $\Delta_a  > 0$, with $\Delta_a  = 0 $ being the borderline case. 
Should the deal be rejected, $R_a(t_1) = R_a(t_0)$ and $R_p(t_1) = R_p(t_0)$, and, consequently, $\Delta_a = 0$. If we denote by $R_{max}$ the total amount of money to be shared and by $R_{share}$ the actual amount money that the proposer has reserved for the accepter, then in the case where the transaction does take place we have 
$R_a(t_1) = R_a(t_0) + R_{share}$ and $R_p(t_1) = R_p(t_0) + (R_{max} - R_{share})$. If
we further assume that $R_a(t_0) = R_p(t_0)$
(i.e. the players start on equal footing, which may very well be the case in the context of the game at least), it follows from Eq.(\ref{perusta}) that the smallest offer that the accepter expects from the proposer is one third of the maximum $R_{max}$, i.e. the condition
\begin{equation}
R_{share} > \frac{R_{max}}{3}
\label{accepted}
\end{equation}
must hold for the proposal to be acceptable. 

Previous literature shows that the minimum offers that people are usually willing to accept are around $30\%$ of a given quantity \cite{kysymys}, in close agreement with the calculation above. Moreover, we note that if the term $R_a(t_1) - R_a(t_0)$ in Eq. (\ref{perusta}) is neglected, then the accepter will never settle for less than half of the total amount.

Next we use Eq.~(\ref{perusta}) to illustrate how the BTH-based approach can be implemented in the context of agent based social simulations. The trick is to generalise this equation to the cases of many players, and with multiple and different kinds of items being exchanged. If there are $N$ players (denoted by $i=1, \ldots, N$), the change of status $\Delta_i$ of the individual $i$ may be written as follows
\begin{eqnarray}
\Delta_i &=& R_i(t_1) - R_i(t_0) \nonumber \\
&& + \sum_{j \neq i} [ R_i(t_1) - R_j(t_1) ] \nonumber \\
&& - \sum_{j \neq i} [ R_i(t_0) - R_j(t_0) ].
\label{glin}
\end{eqnarray}
In the case that players have several ways to measure their status, we may introduce normalisation factors to compare the relative value of the exchanged items, and write the change of status as,
\begin{eqnarray}
\Delta_i &=& \sum_{\alpha} \frac{1}{\mathcal{N}^{\alpha}_i} \bigg\{ R^{\alpha}_i(t_1) - R^{\alpha}_i(t_0) \bigg. \nonumber \\ 
&& + \sum_{j \neq i} \left[ R^{\alpha}_i(t_1) - R^{\alpha}_j(t_1) \right] \nonumber \\ 
&& - \bigg. \sum_{j \neq i } \left[ R^{\alpha}_i(t_0) - R^{\alpha}_j(t_0) \right] \bigg\}
\label{ydelta}
\end{eqnarray}
where the index $\alpha$ runs over the various items determining the status, and $\mathcal{N}^{\alpha}_i$ is the normalisation factor of each item, which may vary from one player to another. The utility function associated with Eq. (\ref{ydelta}) reads then 
\begin{equation}
U^{BTH}_i = \sum_{\alpha} \frac{1}{\mathcal{N}^{\alpha}_i} \bigg\{ R^{\alpha}_i + \sum_{j \neq i} \left[ R^{\alpha}_i - R^{\alpha}_j \right] \bigg\}.
\end{equation}
 
\section{The BTH and agent-based network model}
\label{modelv1}

In this section we present an agent-based model of a ranked social system of $N$ agents, in which the agents exchange their views of the ranking system itself. To each agent $i$ we assign a parameter $a_i$ to describe the rank of the agent, and a state variable $x_i$ to denote the opinion of the agent $i$ of the social value attached to parameter $a$. The social value is then a relative quantity in the minds of the agents, and they value each other in either ascending or descending order according to the ``ranking parameter'', 
we call $a$, and $x_i$ determines which order a given agent $i$ prefers and how strongly. Generally speaking, the sign of $x_i$ represents the chosen order, $-$ for descending and $+$ for ascending order, while its magnitude represents the strength of conviction: with $\lvert x_i \lvert = 0$ the agent can be said to support equality of all the agents irrespective of the ranking parameter, with $\lvert x_i \lvert = 1$ the agent thinks that $a$ should directly define the hierarchy of the society, and for the case $\lvert x_i \lvert < 1$ or $\lvert x_i \lvert > 1$ correspondinly downplaying or emphasizing the significance of $a$, respectively. 
Here, we adopt the maximum value for $\lvert x_i \lvert$ to be $1000$.

To put the relation of the social value and the opinion parameter into more precise terms, we adopt the following expression for the term we call ``ranking pressure'':

\begin{equation}
P_j = \frac{1}{max(a) - min(a)} \sum_k (a_j-a_k),
\end{equation}
where the summation is over the whole network. Now, we define the social value of the agent $i$ in the eyes of agent $j$ as $x_j P_i$, and assume that the agents take into account the views of their neighbours in addition to their own when evaluating their total social value $V_i$. Thus, we write 

\begin{equation}
V_i = x_i P_i + \sum_{k \in m_1(i)} x_k P_i,
\label{totsosval}
\end{equation}
where $m_1(i)$ denotes agents that are one step away from agent $i$. The first term on the right could be considered as the agent's  ``self esteem'' and the second the ``social value'' given to it by its first neighbours.  It should be noted at this point that in defining $V_i$ in terms of $P_i$ we have assumed that $a_i$ does not confer direct advantages or disadvantages for the agents. Therefore, the most natural interpretation for $a_i$ is that it represents the ownership of pure status symbols or Veblen goods, i.e. goods that are only, or mostly, desirable due to their status-enhancing properties, such as luxury items. If $a_i$ would give some advantages or disadvantages for the agents, Eq. (\ref{totsosval}) would have to be revised accordingly. The agents in our simulations attempt to gain as much social value as possible, both in absolute and relative terms, and do this either by changing their opinion variables or adjusting their relations to other agents. The system we use here is purely reactive, with agents reacting to the changes in their social environment in accordance with BTH. The social gain function could then be written as

\begin{eqnarray}
\Delta_i(t_1,t_0)&=& V_{i}(t_1) - V_{i}(t_0) + \sum_{k} (V_{i}(t_1) - V_{k}(t_1)) \nonumber \\
&&  - \sum_{k} (V_{i}(t_0) - V_{k}(t_0)),
\label{altdel}
\end{eqnarray}
where $t_0$ and $t_1$ are the two consecutive time steps. The sign of this function determines the direction of the changes in $x_i$. 

The decision making method employed by the agents is thus a simple hill climbing algorithm: At a given time step $t$, first an agent observes the quantity $\Delta_i(t-1,t-2)$, and then changes its variable $x_i(t-1)$, which leads to a recurrence relation of the form 

\begin{equation}
\label{dm}
x_i(t)=
\left\{
\begin{aligned}
&x_i(t-1)+dx,\;\;\;\; \mathrm{if}\;\;\;\; G_i> 0,\\
&x_i(t-1)-dx,\;\;\;\; \mathrm{otherwise}
\end{aligned}
\right.
\end{equation}
where $G_i=\mathrm{sign} [(\Delta_i (t-1,t-2)\times (x_i(t-2)-x_i(t-1))]$ and $dx$ is a small increment. In the spirit of simulated annealing techniques, the magnitude of the change is larger at the beginning of the dynamics and falls linearly with time to a minimum value, the maximum and minimum values being $dx=0.11$ and $dx=0.01$, respectively, and the time period to reach the minimum is 1000 time steps.

In general, the links between the agents in the social network may change in time for which purpose we use the following rewiring scheme of Ref.~\cite{IKKB2009}.  
The social network of the agents is initially random, but will change periodically, i.e. at every $g$ time steps of the dynamical Eq.~(\ref{dm}). Given the definition of the total social value of an agent in Eq. (\ref{totsosval}), the gain function Eq.(\ref{altdel}) can be used to calculate the loss or gain in total social status when forming or breaking new social bonds. In this study, we take any positive gain as sufficient to justify the rearrangement of social relations between the agents. When agent $i$ considers cutting an existing bond with agent $j$, the gain function has the form
\begin{equation}
\Delta^c_{i,j} = V_j - V_i - (k_i + 2) x_j P_i,
\label{cutgainij}
\end{equation}
where $k_i$ is the current number of neighbours of agent $i$. Similarly, when agent $i$ considers forming a new bond with agent $j$, the social gain function reads
\begin{equation}
\Delta^f_{i,j} = V_i - V_j + (k_i + 2) x_j P_i - x_i P_j.
\label{creategainij}
\end{equation}
 
Since any positive change indicated by the functions above leads to rewiring, it is the sign of these functions that determines whether or not links between agents are broken or created.  For instance, if $\Delta^c_{i,j} > 0$, the link between agents $i$ and $j$ will be cut, and preserved if $\Delta^c_{i,j}  < 0$. In the same vein, a link between agents $i$ and $j$ will be created if $\Delta^f_{i,j} > 0$, and not created otherwise. It should be noted that when forming links the opinions of both agents are taken into account: The relation formation only succeeds if $\Delta^f_{i,j}$ and $\Delta^f_{j,i}$ are both positive. The agents will form all the relationships they can in a rewiring cycle.

\section{Numerical results}
\label{nc}

The numerical simulations of the models of social system described in the previous sections are performed as follows. First, the initial state of the system is set at random with the agents given a relatively small initial opinion $x_i$ between $-1$ and $1$, a ranking parameter $a_i$ between $0$ and maximum value of $100.0$ and initial connections to other agents, with initial average degree of $5$. The opinion and ranking parameters are chosen using a random number generator, which returns a flat distribution. The dynamics are then run for $200000$ time steps, which is, according to our test runs, sufficient for the general structure of the network to settle. However, the dynamics of the opinion variables do not have a set stopping point, so they may experience fluctuations even when such fluctuations do not  have an effect on the network structure anymore. 

To obtain reliable statistics, the same simulations are repeated $100$ times with random initial values, and averages are calculated from these repeated tests for the quantities under study. The rewiring timescale $g$ is fixed to $100$ in our simulations, since this value lies in the range where communities are formed in the opinion formation model of Ref. \cite{IKKB2009}. The main parameter whose effect is studied here is the number of simulated agents, $N$. 

The main objective of this research is to study the structure of the social networks created under BTH assumption in the case of a rigid ranking system, which we perform using the model explained in Section \ref{modelv1}. The most interesting properties of the system are then associated with assortativity, or the tendency of agents with high degrees connecting to other highly connected agents, and homophily, or the inclination of similar agents forming connections between each other. In the context of this study, homophily refers to agents with similar ranking parameters forming connections with each other. 

The averaged numerical results extracted from the simulations consist then of the standard network properties, i.e. degree $\langle k \rangle$ the shortest path $\langle L \rangle$, the average clustering coefficient $\langle C \rangle$, the mean number of second neighbours $\langle n^{(2)} \rangle$, susceptibility $\langle s \rangle$ and average assortativity coefficient $\langle r_a \rangle$, and a homophily coefficient $\langle r_h \rangle$. Susceptibility here refers to average cluster size, which is calculated as the second moment of the number of $s$ sized clusters, $n_s$:
\begin{equation}
\langle s \rangle = \frac{\sum_s n_s s^2 }{\sum_s n_s s}.
\label{suscep}
\end{equation}
As customary in percolation theory, the largest connected component of the network is not counted in calculating $s$. For the assortativity coefficient we use the definition given in \cite{PPZ}, and the homophily coefficient is defined using Pearson's product moment coefficient, which measures the goodness of a linear fit to a given data. For a sample it can be defined as 
\begin{equation}
r_h = \frac{\sum_i^M (v_i - \overline{v}) (w_i - \overline{w})}{\sqrt{\sum_i^M (v_i - \overline{v})^2}\sqrt{\sum_i^M (w_i - \overline{w})^2}},
\label{homophilypar1}
\end{equation}
where $v$ and $w$ are vectors containing the value parameters of agents linked by link $i$, $\overline{v}$ and $\overline{w}$ are the mean values of these vectors, respectively, and $M$ is the total number of links. More specifically, if agents $\alpha$ and $\beta$ are connected by link $i$, then $v_i = a_{\alpha}$ and $w_i = a_{\beta}$. The links are indexed as follows: the links involving the first agent are given the first indices, then follow the links involving the second agent but not the first, and so on, without repeating links that have already been indexed. It should be noted, however, that $r_h$ only measures linear correlation between the ranking parameters of linked agents, it does not indicate how steep these trends are. To check whether the system is truly homophilic, then, one needs to make a linear fit to the data: the closer the obtained linear coefficients are to $1$, the greater the homophily.

\begin{figure}[h]
\begin{center}
\epsfxsize = 0.85\columnwidth \epsffile{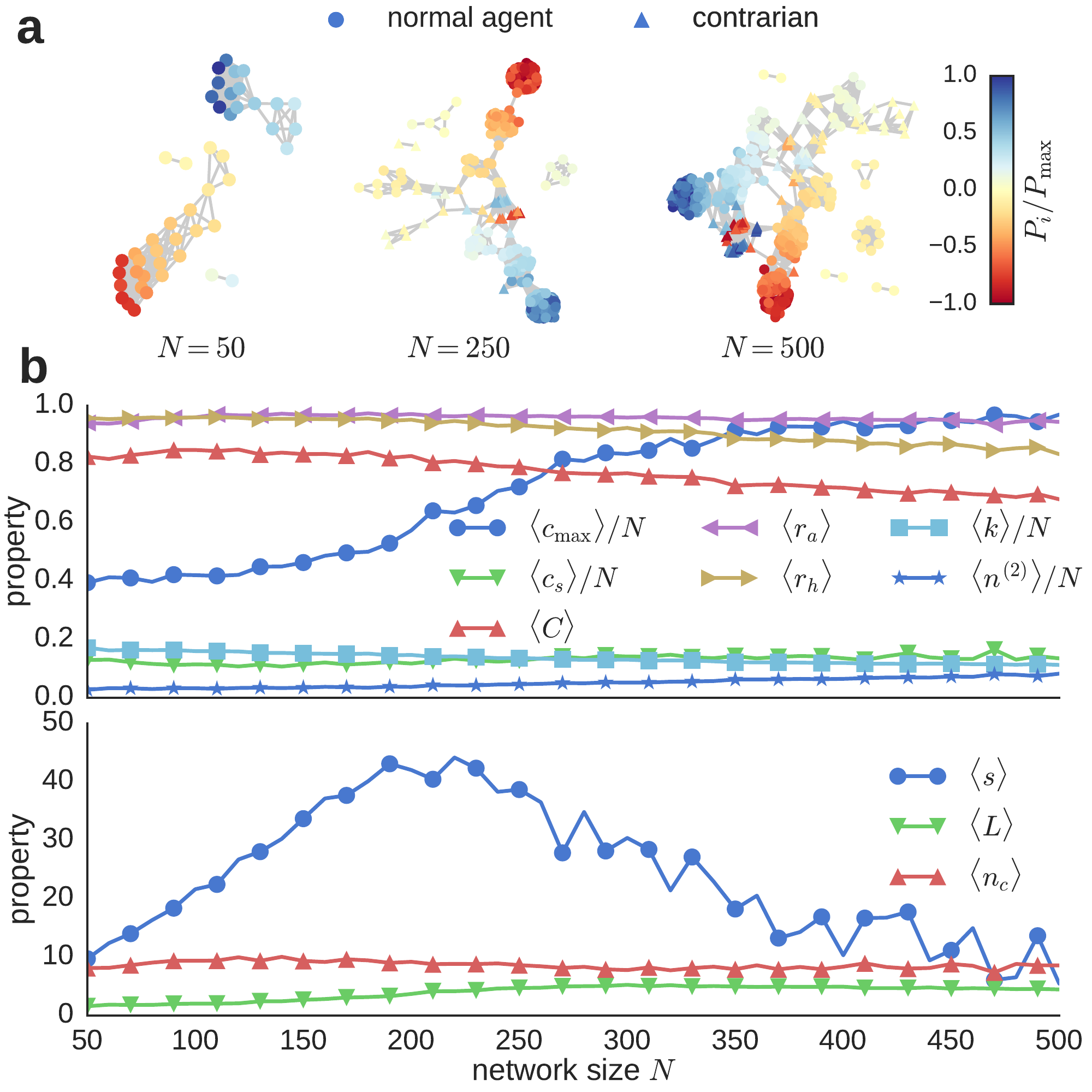}
\caption{(a) Graphs showing examples of the network structure for three different sizes, $N$, where the colour represents the value of the agent's normalized ranking pressure, $P_i/P_{max}$. Observe communities of tightly bound agents of the same colour. The circular vertices represent agents whose opinion variable $x_i$ and ranking pressure $P_i$ have the same sign, while the triangular vertices represent agents having opposite signs. (b) Network properties, such as  degree $\langle k \rangle$, the shortest path $\langle L \rangle$, the average clustering coefficient $\langle C \rangle$, the mean number of second neighbours $\langle n^{(2)} \rangle$, susceptibility $\langle s \rangle$,  assortativity coefficient $\langle r_a \rangle$, and homophily coefficient $\langle r_h \rangle$  as averages over 100 realizations for the model as a function of the population size, ranging from $50$ to $500$.}
\label{fig:NP100r}
\end{center}
\end{figure}

The average network properties of the system, with graphs illustrating the behaviour of the system are shown in Fig.~\ref{fig:NP100r} as functions of the population size, which is varied between $50$ and $500$. The main observations that can be made about the graphs in Fig.\ref{fig:NP100r}(a) are that at lower population levels they show a tendency of breaking apart into many subcomponents of different sizes, while for larger population sizes they tend to consist of a single large component and possibly some smaller separate clusters. A noteworthy fact about these clusters is that they consist of agents with similar values of the ranking pressure $P_i$, which means that the network exhibits homophily in this case. The largest clusters are found at extreme values, and they become smaller when one approaches $0$, which also corresponds to average ranking parameters. 

In the high population case the picture becomes more complicated due to the emergence of clusters that contain agents with opposing opinions as well. These new clusters tend to be less connected than the previously described homogeneous ones, and they tend to connect to the large subgraphs, thus forming a single giant graph. 
A closer look reveals that these agents generally have opinion variables and ranking pressures with opposite signs and are depicted as triangles in Fig. \ref{fig:NP100r} and named `contrarians'' from now on.
 
\begin{figure}[h]
\begin{center}
\epsfxsize = 0.85\columnwidth \epsffile{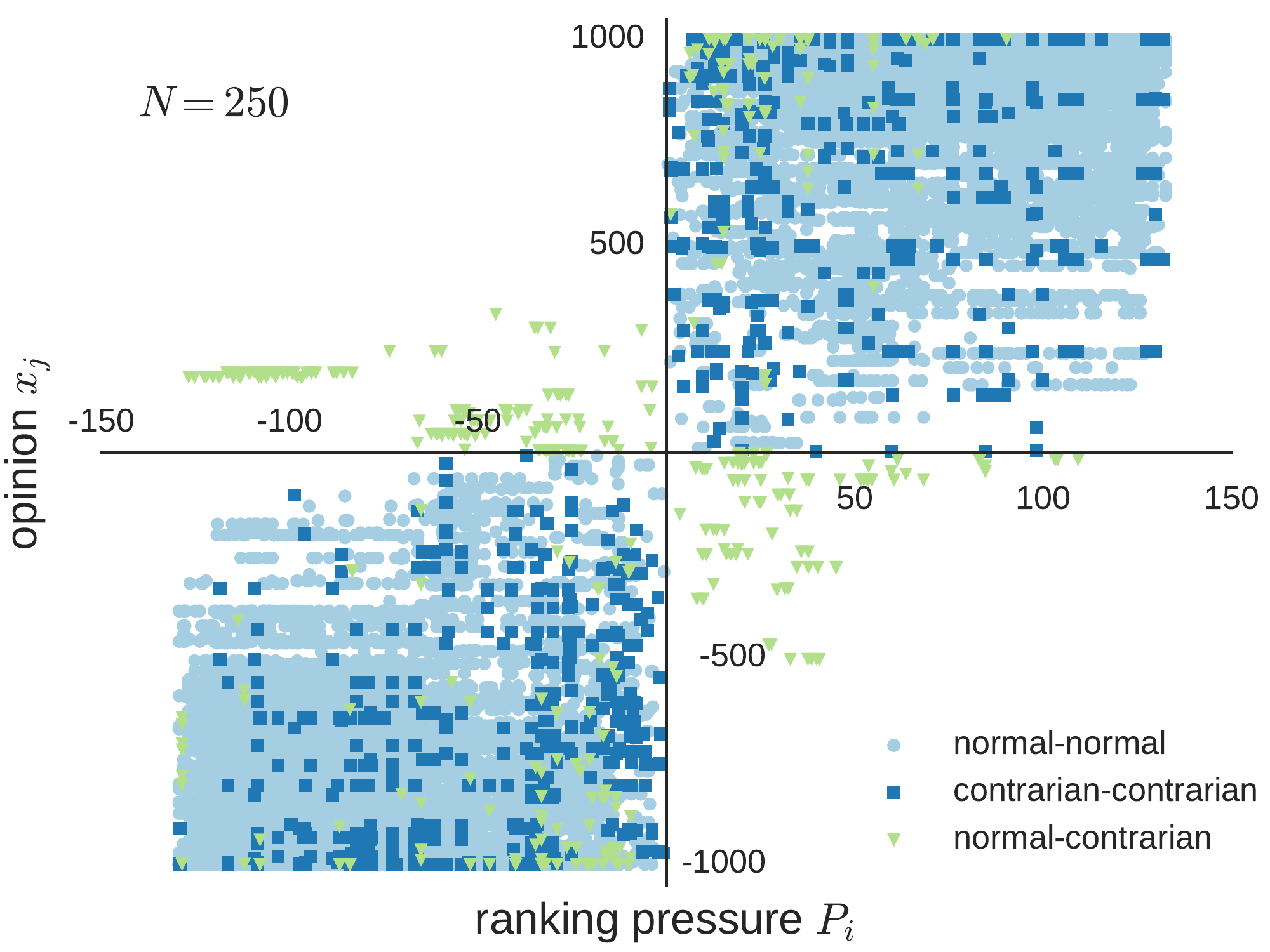}
\caption{Scatterplot for ranking pressure $P_i$ of agent $i$ vs. the opinion variable $x_j$ of agent $j$, taken from 10 realisations for networks with $N=250$ agents. Observe that in most cases $x_j$ and $P_i$ have the same sign (especially for normal-normal and contrarian-contrarian links), while only in case of normal - contrarian links they can have opposite signs. }
\label{fig:1bis}
\end{center}
\end{figure}

\begin{figure}[ht!]
\begin{center}
\epsfxsize = 0.85\columnwidth \epsffile{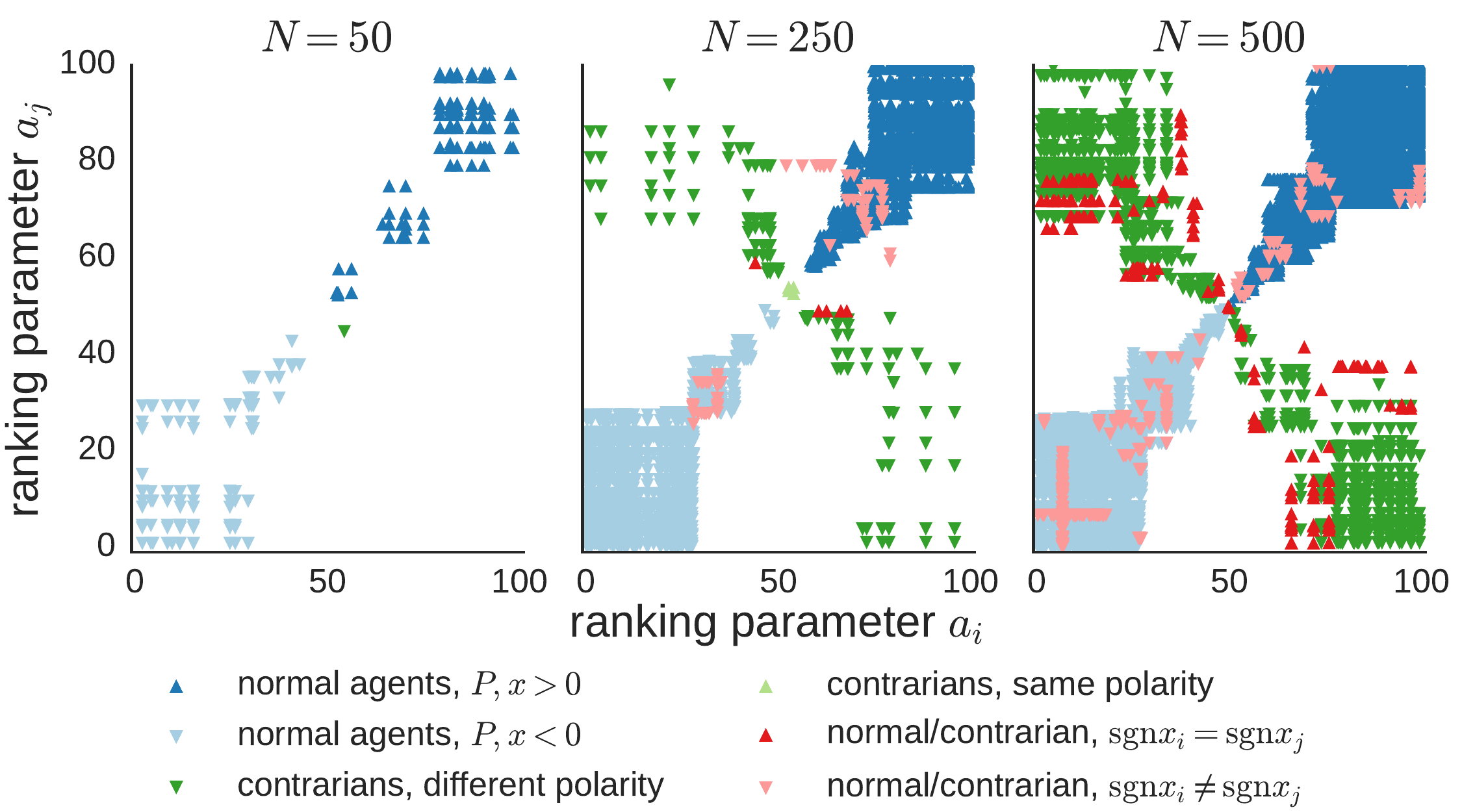}\\  
\caption{A scatterplot of the ranking parameters $a_i$ vs. $a_j$ of linked agents $i$ and $j$ for one realisation of networks of different sizes. The colour code:  dark blue for connections between normal agents with positive ranking pressure and opinion variables,  light blue for connections between normal agents with negative ranking pressure and opinion variables,  dark green for connections between contrarian agents of different "polarity",  light green for connections between contrarian agents of same "polarity" (these are almost non-existent),  light red for connections between normal and contrarian agents with different signs of the opinion variable, and  dark red for connections between normal and contrarian agents with same signs of the opinion variable.}
\label{fig:scatter2}
\end{center}
\end{figure}

A naive analysis would indicate that the agents with positive ranking pressures should always support the ascending hierarchy, and the agents with negative ranking pressures should always support the descending hierarchy. However, the contrarian agents exhibit opposite preferences. The reason why this behaviour is status-wise profitable can be found by looking into the connections of the contrarian agents, details of which are shown in Figs. \ref{fig:1bis} and \ref{fig:scatter2}. As it turns out, most of the connections they form are to other similar agents but with opposite ``polarity'' to theirs, i.e. the contrarian agent with negative ranking pressure forms connections mostly with contrarian agents of positive ranking pressure, and vice versa. 
 
 An important quantity in the model is the social value $V_i$, which is a product of the opinion and ranking pressure,  as it is seen in Eq.~\ref{totsosval}. 
If $x_i$ and $P_i$ have opposite signs then the ``self esteem'' part of $V_i$ is negative, which in itself does not mean that the agent cannot develop a contrarian opinion, since the second sum in the equation could be positive because it depends on the opinions of the neighbours $j$. This allows the contrarian to be able to make connections with agents of the same or opposite ranking pressure. Additionally, by looking at all connections among contrarians we find that they mostly have $P_i$ of the same sign as $x_j$, as illustrated in the example of Fig.~\ref{fig:1bis}.
 
This situation can be status-wise beneficial to all parties involved, since the small penalty to an agent's self-esteem is more than compensated by the respect that the agent will gain in this case from other agents. The fact that the agents could find this strategy using as primitive an intelligence setup as hill climbing is astounding. Another interesting thing about the contrarians is that they appear mostly as connections between clusters that are defined as communities of ``normal'' agents.

The various kinds of behaviour exhibited by the social networks have a marked effect on the network properties also shown in Fig.~\ref{fig:NP100r}(b). The most obvious is the gradually rising normalised maximum cluster size ($\langle c_{\mathrm{max}}\rangle/N$), which is about 40\% of total population size for $N = 50$, and over 95\% for $N = 500$. From the figure it seems that the maximum cluster size reaches 50\% of the population size for approximately $N = 180$, after which point we may assume that the contrarian behavioural patterns start to become progressively more pronounced. 
 
 The susceptibility $(\langle s\rangle)$ at first rises pretty much linearly, which is not too surprising because of the tendency of network to break into smaller subgraphs at low population sizes. However, once the population size reaches about $200$, the susceptibility starts to decay, most likely due to the main component of the network becoming more prominent, with a decay pattern that is almost piecewise linear itself, apart from fairly large fluctuations. 
 
 While fairly high throughout, the homophily $(\langle r_h\rangle)$ and clustering coefficients $(\langle C\rangle)$ gradually fall as functions of population size, almost certainly due to the proliferation of contrarians. There are no great changes in other network properties, as in the average assortativity coefficient ($\langle r_{a}\rangle$), although some faint systematic tendencies can be discerned, a slight rising of the average path length ($\langle L\rangle$), as well as slightly decreasing average number of clusters ($\langle n_{c}\rangle$). The rest of the properties per agent are nearly constant, a slight rise of the  average number number of second neighbours ($\langle n^{(2)}\rangle/N$), and a just perceptible decrease of the average cluster size ($\langle c_{s}\rangle/N$) and average degree ($\langle k\rangle/N$).

 A way to illustrate the homophily of the system is to make a scatterplot of the ranking parameters of linked agents. As it is seen in Fig.\ref{fig:scatter2}, the correlation turns out to be very homophilous, as the ranking parameters of linked agents correspond very closely to one other. The emergence of the contrarians is also clearly seen: for $N = 50$, the percentage of contrarians in 10 realisations is 5.4\%, while for $N = 250$ is 13.8\%, for $N=500$ is 24.9\% and for one realisation in in a network of 1000 agents is 38.5\%. There is also a clear decreasing linear trend due to the contrarians. 

The rising trend is without doubt caused by the normal agents, who tend to associate with agents of similar rank, and the decreasing trend is likewise due to the contrarians. Both trends have a similar tendency to form square-like patterns along the diagonals, with each ``square'' corresponding to some of the many visible communities of the graphs. 

As explained above,  it is necessary to check whether the correspondence of the rankings is truly homophilic. In Table~\ref{tab:1} we show the value of the homophily coefficient of normal and contrarian agents for networks of various sizes. The data were taken from 10 different numerical realisations in each case. Observe that normal agents have values very near one, and contrarians are around one half. Also in the table we show  the slope of the regression of $a_i$ vs. $a_j$. Normal agents are very close to one, indicating high degree of homophily, and the contrarians are negative and around 0.6, indicating that they mostly form connection with agents that have opposite signs of the ranking pressure, and are much less homophilic. 

\begin{table}[ht!]  
\begin{tabular}{l c c c c c c c c} 
& \multicolumn{4}{c}{Normals} 
& \multicolumn{3}{c}{Contrarians} \\
\hline \hline
  & N=50 & N=250 &N=500 & \vline & N=50 & N=250 &N=500\\
 \hline
Slope & 0.955 & 0.968 & 0.963 & \vline & -0.606 & -0.607 & -0.707\\
$r_h$ & 0.959 & 0.963 & 0.959 & \vline & -0.588 & -0.636 & -0.731\\ 
\hline
\end{tabular}
\caption{Homophily measurement from 10 numerical realisations of networks with different sizes.} \label{tab:1} 
\end{table}

\begin{figure}[ht!]
\begin{center}
\epsfxsize = 0.85\columnwidth \epsffile{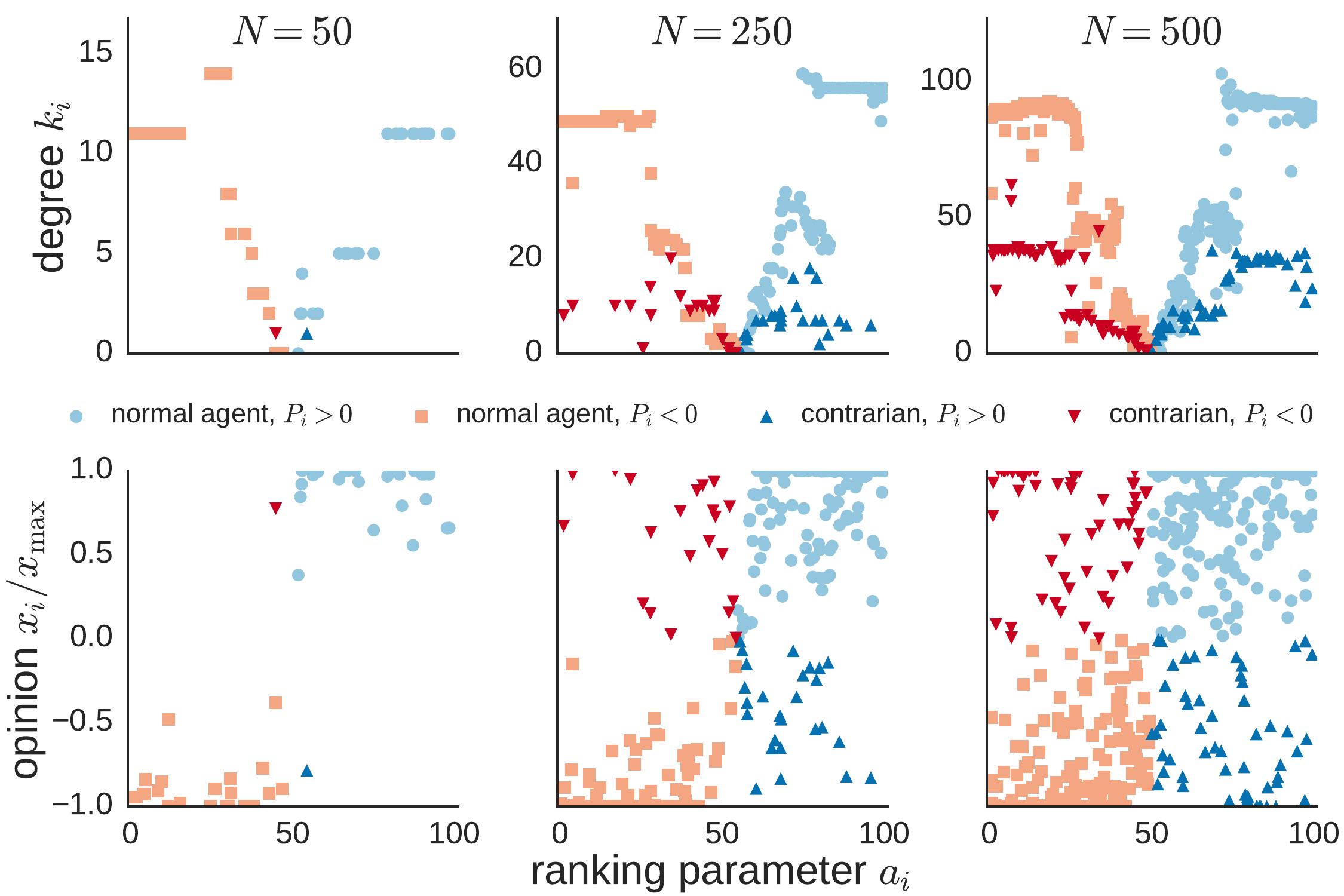}\\ 
\caption{ The degree $k_i$ and the final state of the opinion variable $x_i/x_{max}$ as  a function  of the ranking parameter $a_i$ of agent $i$ . The colour code is as follows: light blue for normal agents with positive ranking pressure, light red for normal agents with negative ranking pressure, dark blue for contrarian agents with positive ranking pressure, and dark red for contrarian agents with negative ranking pressure.}
\label{fig:scatter3}
\end{center} 
\end{figure}

In Fig.~\ref{fig:scatter3} we show the degree and the final state of the opinion variable as  functions  of the ranking parameter of the agents. One can see that the agents with relatively large or low values for the ranking parameter seem to have more neighbours than the agents with the ranking parameters close to the average rank. The communities are also visible in this figure in the form of plateaus at progressively more extreme values of the ranking parameter. The appearance of contrarians is also very clearly visibly in this figure.

The lower row of panels in Fig.\ref{fig:scatter3} shows the correlation of agents' ranking parameters and opinion variables. As expected, most of the agents below the average ranking parameters have negative opinion variables, and similarly, the agents with the above average ranking parameters have positive opinion variables, at least for small $N$. 
There are some exceptions to this rule. The contrarian agents are seen clearly in the figure as agents whose opinion variables have opposing signs to agents of similar ranking. On the other hand, when they are present, there are often such normal agents that have more egalitarian views, i.e. less extreme opinions.

\begin{figure}[ht!]
\begin{center}
\epsfxsize = 0.85\columnwidth \epsffile{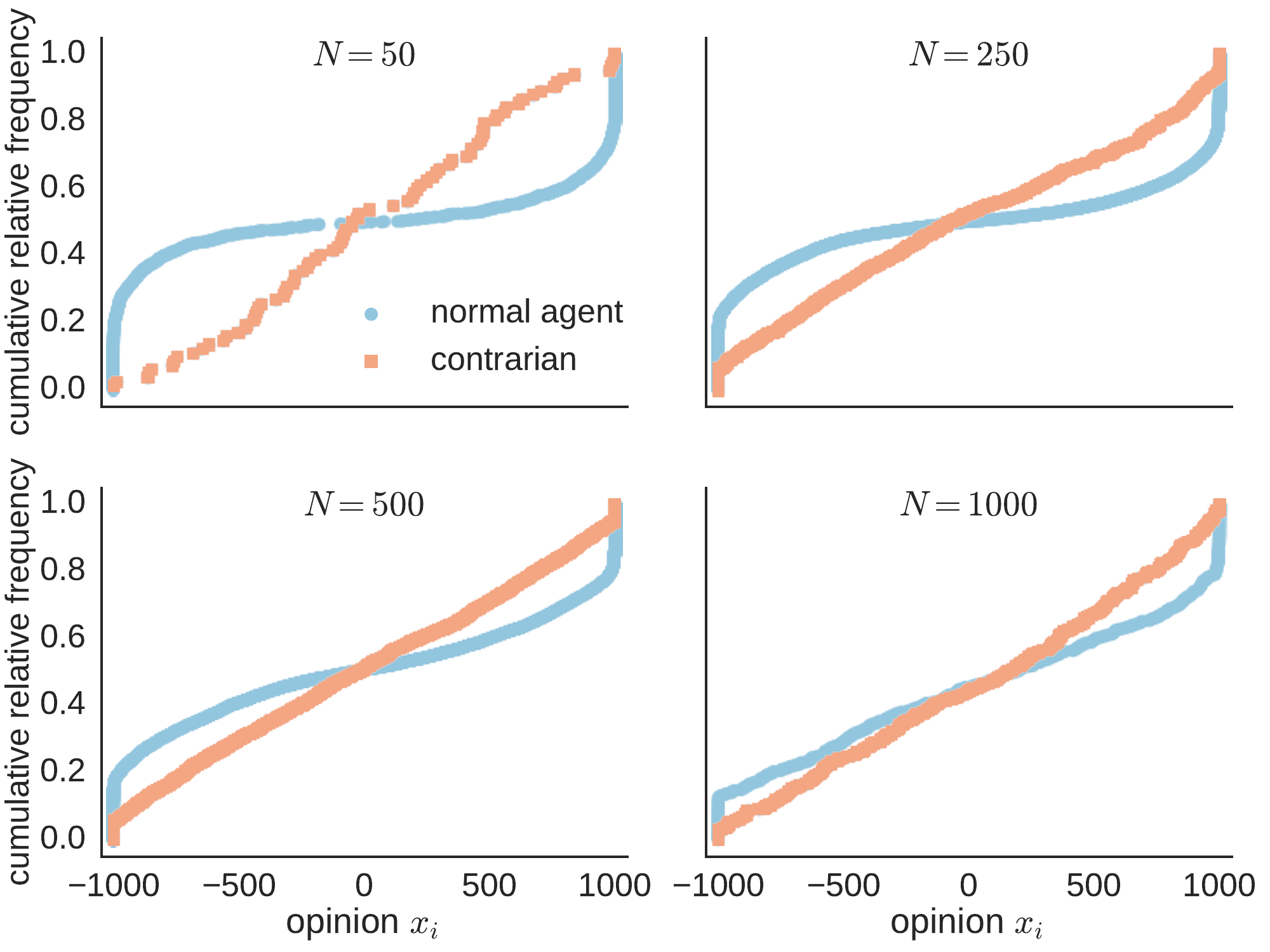}\\ 
\caption{Empirical distributions of opinions for 10 realisations on networks of sizes 50, 250, and 500 and for single realisation for the system size of 1000 agents, where blue and orange symbols stand for normal and contrarian agents, respectively. }
\label{fig:crf}
\end{center}
\end{figure}

There is a clear tendency for the agents to attain the most extreme values for opinion variables in either case, although this tendency is smaller as the size of the network increases. A clear visualisation of this phenomenon is depicted in Fig.~\ref{fig:crf} where one can see that the cumulative relative frequency of values of $x$ is almost vertical at the extremes for normal agents. The proportion of normal agents  with $|x|>980$ is 51.6, 47.4, 39.5 and 34.3\%, for $N=$ 50, 250, 500 and 1000, respectively. The picture is very different for contrarians, their distribution tends to be uniform independent of the network size, except at the extreme values where the percentage of agents with $|x|>980$ is 4.8, 10.4, 11.6 and 6.4\%, for $N=$ 50, 250, 500 and 1000, respectively.

\begin{figure}[ht!]
\begin{center}
\epsfxsize = 0.85\columnwidth \epsffile{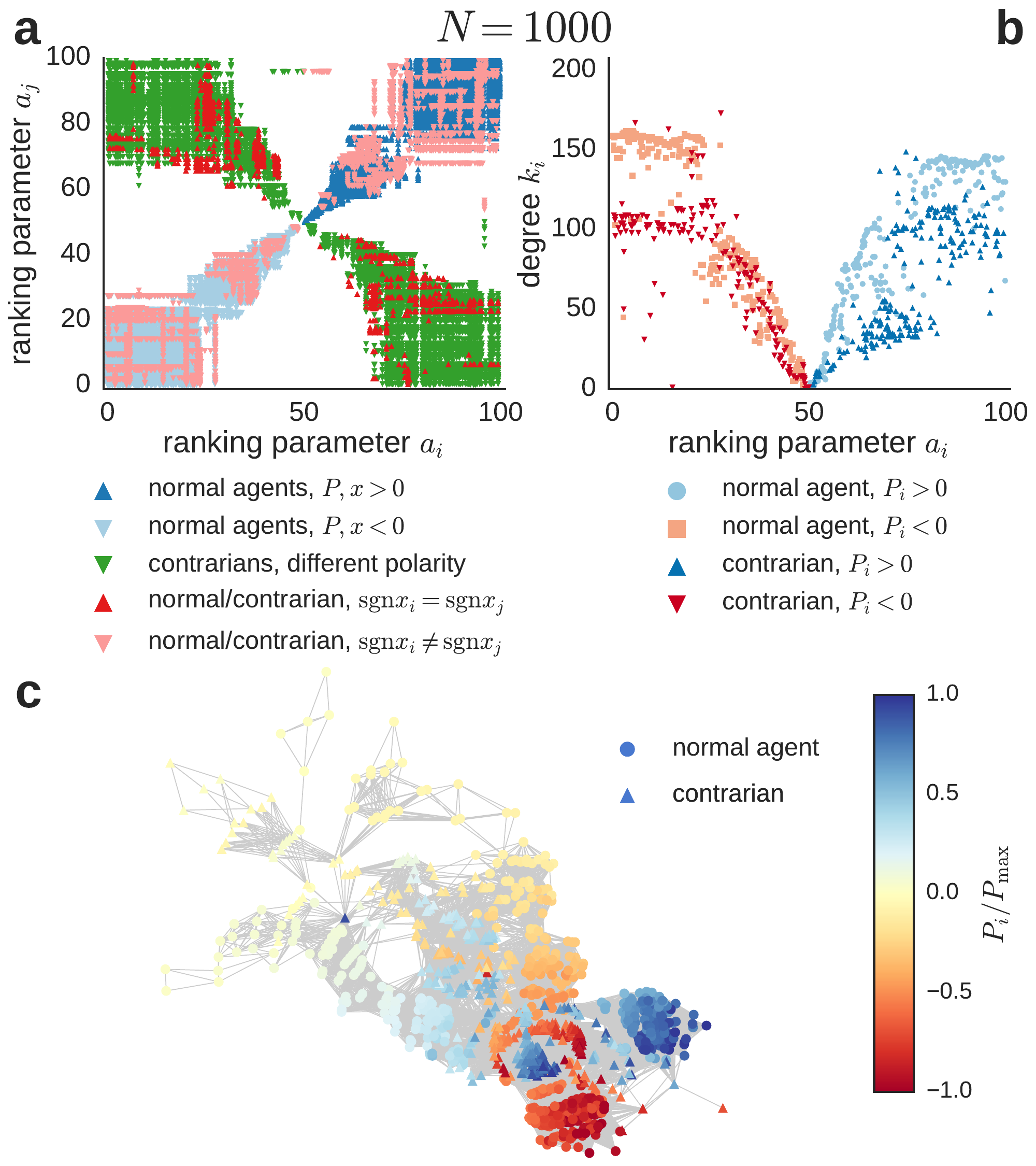}\\  
\caption{The rank-rank (a) and rank-degree correlations (b) for a single simulation with $N = 1000$ agents, along with the final configuration of the network (c); the colour codes and symbols are the same as in Figs  \ref{fig:NP100r} (c), \ref{fig:scatter2} (a), and \ref{fig:scatter3} (b).}
\label{fig:p1000}
\end{center}
\end{figure}

As simulations with over $500$ agents are very time-consuming, we have not tried to obtain results with large statistics for these cases. We did, however, run some singular simulations with very high agent numbers, to see whether the patterns observed above would hold even there. Fig. \ref{fig:p1000} shows the resultant graph, rank-rank, and rank-degree correlation scatterplots for a simulation with $1000$ agents. As we can see, the contrarians have become more numerous and have ever more extreme ranking parameters, as could be expected from the earlier results. Another interesting feature that emerges from the analysis of 10 numerical realisations is that the degree distribution is bimodal for normal agents,  that is, there is a large number of agents with low or high degree and very few around the mean degree value. The range of degrees and the median of the distribution increase with the size of the network. 

\begin{figure}[ht!]
\begin{center}
\epsfxsize = 0.85\columnwidth \epsffile{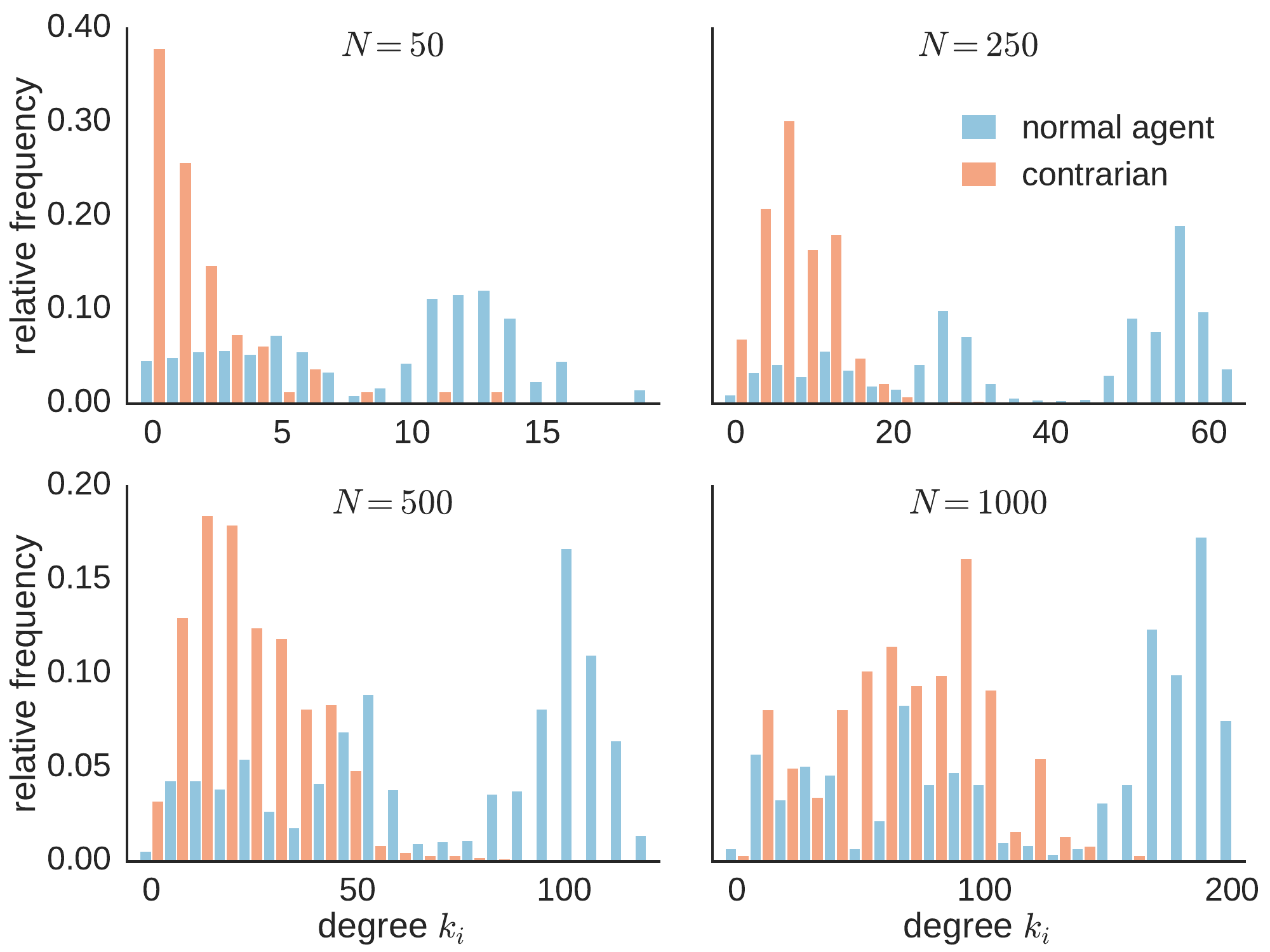}\\  
\caption{Histogram showing the degree distribution of agents for 10 realisations on networks of sizes 50, 250, and 500 and for single realisation for the system size of 1000 agents.}
\label{fig6}
\end{center}
\end{figure}

For contrarians the picture is different. their degree  distribution is unimodal (although not normal) and their median is smaller. Furthermore, their degree range is about half the one for normal agents. This is clearly seen in Fig.~\ref{fig6}, were we plot the relative frequency as a function of the degree $(k)$. Observe that the proportion of isolated agents ($k=0$) is larger for the contrarians although this tendency diminishes as the networks become larger.

\section{Approximative analysis of the numerical results}
\label{meanfield}

In order to better understand the behaviour of the model we need to perform a thorough analysis of its mechanics. A convenient starting point for this effort is studying the tendency of the agents with extreme ranking parameters to form large fully connected communities, while the agents of more average rank form small fully connected communities. Just from considering the nature of the BTH one might formulate the hypothesis that this may be because of agents with average rankings not forming links with agents with extreme rankings due to it not being status-wise beneficial. If an agent sees that the other agent would gain more in having a link between them, that link will probably not be created by the agent. By inspecting the rewiring rules Eqs.(\ref{cutgainij}) and (\ref{creategainij}) this hypothesis can be verified. Without loss of generality, we can limit our investigation to the case of two agents with positive ranking pressures and opinion variables considering forming a link, as the case with agents with negative ranking pressures and opinion variables follows from a similar line of reasoning. 
From Eq.~(\ref{creategainij}) one finds that if agent $i$ considers forming a link with agent $j$, the ranking pressures and the opinion variables of the agents and their neighbours must satisfy the condition 
\begin{equation}
\frac{P_i}{P_j} > \frac{ x_j + \sum_{k \in m_1(j)} x_k +x_i }{(k_i + 2)x_j + x_i + \sum_{k \in m_1(i)} x_k}
\label{expla1}
\end{equation}
for the potential link to be acceptable to agent $i$. If we assume, for simplicity, that the opinion variables of the agents and their neighbours have about the same value, then condition~(\ref{expla1}) simplifies to
\begin{equation}
\frac{P_i}{P_j} > \frac{1}{2} \frac{ k_j + 1 }{k_i + 3/2},
\label{expla2}
\end{equation}
which in turn simplifies to 
\begin{equation}
\frac{P_i}{P_j} > \frac{1}{2},
\label{expla3}
\end{equation}
if we further assume that $k_i,k_j \gg 1$ and $k_i \approx k_j$, which is reasonable considering that most of the agents at least have large number of neighbours and tend to belong to almost fully connected communities. Here the implicit assumption is that the agents $i$ and $j$ would belong to the same block, if they formed a link. Written in terms of ranking parameters (\ref{expla3}) becomes 
\begin{equation}
a_i > \frac{1}{2} \left( a_j + \langle a \rangle \right),
\label{expla4}
\end{equation}
where $\langle a \rangle = N^{-1} \sum_k a_k $ is the average of the ranking parameters. From (\ref{expla4}) it directly follows that if agent $j$ has the maximum allowed ranking parameter, $a_j = a_{max}$, the link between the agents $i$ and $j$ will only be formed if $a_i > 0.75 a_{max}$, since $\langle a \rangle \approx 0.5 a_{max}$. Similarly, if $a_j = 0.75 a_{max}$, the two agents will only bond if $a_i > 0.625 a_{max}$, and if $a_j = 0.625 a_{max}$, only if $a_i > 0.5625 a_{max}$, and with every iteration the range of possible ranking parameters of agent $i$ shrinks. This pattern is remarkably apparent in the rank-rank correlation Figs.~\ref{fig:scatter2} and \ref{fig:p1000}, in which the emergence of the contrarians becomes increasingly clear as the population numbers are increased. For $a_j = 0.5 a_{max}$ we find that $a_i $ must also be $0.5 a_{max}$ for a link to be formed, which explains the tendency of the agents with average ranking parameters to have so few neighbours, as is seen in Figs.\ref{fig:scatter2} and \ref{fig:p1000}.

To understand the emergence of the contrarians one must analyze the dynamics of the opinion variables, encapsulated in Eqs.(\ref{altdel}) and (\ref{dm}). Naively thinking, one would expect agents to always choose the orientation of their opinion parameters according to their ranking pressures. This means that agents with positive ranking pressures would prefer positive opinion variables, and similarly agents with negative ranking pressures would prefer negative opinion variables. The first hypothesis as to how some agents would choose to go against these logical positions is related to the competitive nature of the model's social interactions. It may be that some agents cannot compete in this setting and, therefore, choose to use contrarian strategies instead.

 Let us approach this question the same way as above, focusing on an agent with positive ranking pressure and initially positive opinion variable, connected to other similar agents in the way revealed in Section \ref{nc}. Let us assume that, initially, all the agents with positive ranking pressure increase their opinion variables and, conversely, that all the agents with negative ranking pressure decrease their opinion variables in a simulation step. In this case the social gain function of agent $i$ is 
\begin{eqnarray}
\Delta_i(t_1,t_0) &=& dx (N +1) ( k_i + 1) P_i - dx \sum_{j \in P^+}  (k_j + 1) P_j \nonumber \\
&& + dx \sum_{j \in P^-}  (k_j + 1) P_j ,
\label{cgfu}
\end{eqnarray}
where $P^+$ is the set of agents with positive ranking pressure and $P^-$ the set of agents with negative ranking pressure. Now,$\Delta_i(t_1,t_0)$  needs to be positive if agent $i$ is to continue raising its opinion variable. 
This yields a condition of the form
\begin{equation}
(k_i + 1)P_i > \frac{1}{N + 1} \left( \sum_{j \in P^+}  (k_j + 1) P_j - \sum_{j \in P^-}  (k_j + 1) P_j \right).
\end{equation}
If we denote the number of agents in $P^+$ and $P^-$ with $N^+$ and $N^-$, respectively, we can write this formula in terms of averages over $P^+$ and $P^-$ as follows:
\begin{equation}
(k_i + 1)P_i > \frac{1}{N + 1} \left(N^+ \langle  (k + 1) P \rangle^+ - N^- \langle (k + 1) P \rangle^- \right),
\label{cghelp}
\end{equation}
where $\langle \rangle^+$ denotes average over $P^+$, and, likewise, $\langle \rangle^-$ denotes average over $P^-$. If we assume $N \gg 1$ and that the situation is symmetric, i.e. 
\begin{equation}
N^+ \langle  (k + 1) P \rangle^+ \approx -N^- \langle (k + 1) P \rangle^-, 
\end{equation}
which further simplifies to 
\begin{equation}
(k_i + 1)P_i > \langle  (k + 1) P \rangle^+,
\label{cgsimple}
\end{equation}
since $N = N^+ + N^-$ and $ N^+ \approx N^-$. 

Condition (\ref{cgsimple}) basically states that revising opinion parameter upwards, which is the ``natural'' direction for an agent in $P^+$, is only status-wise beneficial for the agent if the product of the agent's degree (plus one) and the ranking pressure are above the average of the same product over the whole group $P^+$. We can see from this that the agents with low ranking pressures and the amounts of connections fare badly under the condition (\ref{cgsimple}). As observed earlier, in the simulations the agents with average ranking parameters, who correspondingly have low ranking pressures, are also the ones with least connections, which accordingly means that they are most likely to become contrarians, at least with regard to (\ref{cgsimple}). 

But what about the observed phenomenon of more and more of agents with higher and higher ranking parameters (and thus ranking pressures) to become contrarians as the total numbers of simulated agents increases?. The answer to this question can be found by substituting the numbers obtained for the sizes of the different communities to the right hand side of inequality (\ref{cgsimple}), along with the average ranking parameters for these communities, as shown in Table~\ref{t:1}. 

\begin{table}[ht!]
\begin{tabular}{clcl}
$n$
& \multicolumn{1}{c}{$M_n$} 
& \multicolumn{2}{c}{$\langle a \rangle_n$} \\ 
\hline \hline
$1$ \vline & $0.25 N$  & \vline   &  $0.875 a_{max}$ \\ \hline
$2$ \vline & $0.125 N$ & \vline   &   $0.6875 a_{max}$ \\ \hline
$3$ \vline  & $0.0625 N$ & \vline   &   $ 0.59375 a_{max}$ \\ \hline
$4$ \vline & $0.03125 N$ & \vline   &   $ 0.546875 a_{max}$ \\ \hline
\end{tabular}
\caption{The approximate number of members and the average ranking parameters of the 4 largest communities in $P^+$.}
\label{t:1}
\end{table}

From condition(\ref{expla4}) it follows that the largest community in $P^+$ comprises of those agents with ranking parameters over $0.75 a_{max}$, which means that the community will have approximately $0.25 N$ members, when one takes into account the fact that the ranking parameters are uniformly distributed. The second largest community, likewise, consists of those agents with ranking parameters between $0.625 a_{max}$ and $ 0.75 a_{max}$, and has about $0.125 N$ members. The nth ($n > 1$) largest group will have ranking parameters between $a_{max} (1 - \sum^{n + 1}_{i = 2} 2^{-i})$ and $a_{max} (1 - \sum^{i + 2}_{i = 2} 2^{-n})$, and have $2^{-(n + 1)} N$ members. While ranking parameters naturally vary from agent to agent within the communities, the average value 
of the ranking parameters of each group $\langle a \rangle_n$ falls approximately to the middle point of each ranking range due to the uniform distribution of the parameters: 
\begin{equation}
\langle a \rangle_n = a_{max} \left( 1 - ( \sum^{n + 1}_{i = 2} 2^{-i} + 2^{-(n + 1)}) \right).
\end{equation}

From Figs.~\ref{fig:scatter2} and~\ref{fig:p1000} we see that only a maximum of four to five of these communities exist in practice at any one time, so we limit our approximation to these groups. By assuming the groups to be fully connected, as they seem to be in the graphs, and approximating  sums of the products of the degrees and ranking parameters with the products of their average values we get
\begin{equation}
\langle  (k + 1) P \rangle^+ \approx 2 \sum^4_{n = 1} M^2_n \left( \langle a \rangle_n - \frac{ a_{max} }{2} \right),
\label{cgcomplex}
\end{equation}
where $\langle \rangle_n$ denotes the average ranking parameter and $M_n$ the number of members of the $n$th largest group. Substituting the values given in Table~\ref{t:1} we get 
\begin{equation}
\langle  (k + 1) P \rangle^+ \approx 0.05 N^2 a_{max},
\end{equation}
which in turn can be inserted into (\ref{cgsimple}): 
\begin{equation}
(k_i + 1)P_i > 0.05 N^2 a_{max}.
\label{cgsimple2}
\end{equation}
Finally, using the definition of $P_i$ the condition (\ref{cgsimple2}) can be written in the form 
\begin{equation}
\alpha_i  > \frac{1}{2} + 0.05 \frac{N}{k_i + 1} ,
\label{cgsimple3}
\end{equation}
where $\alpha_i = a_i/a_{max}$. 

From condition~(\ref{cgsimple3}) we can see that the probability of agent $i$ following the conventional wisdom diminishes with rising $N$ and decreasing $k_i$, which is what we saw happening in our simulations judging from the results shown in the previous Section. To take an example, for the $N = 500$ case the largest community of agents with positive ranking pressures comprises of about $125$ agents. This means, according to the inequality (\ref{cgsimple3}), that the agents with $\alpha_i > 0.7$ could definitively be expected to always choose to have positive opinion variables. 

From Fig.~\ref{fig:scatter2} we can tell that the real threshold is closer to $\alpha_i > 0.8$, which is, however, in remarkably good agreement with the approximate value of $0.7$ when one takes into the consideration the fact that the appearance of the contrarians themselves was not taken into account in the derivation of  (\ref{cgsimple3}), and that for   $N = 500$ they are already very prominent. If one were to derive the condition equivalent to (\ref{cgsimple3}) with contrarian strategies taken into account, one would need to consider the effect that the contrarians have on their neighbours' total social value $V$. Thus, condition~(\ref{cgsimple3}) will most likely not hold for networks with larger $N$.

The last question we need to address as regards to the contrarians is the fact that they are often embedded in the groups of normal agents. So why would it be status-wise beneficial for an agent using normal strategy to retain, let alone form, a link with a contrarian agent?

 Let us consider a situation where an agent pursues contrarian strategies in a group of normal agents. Returning to inequality~(\ref{expla1}), we see that it is acceptable for a normal agent $i$ with $P_i > 0$ and maximal $x_i = x_{max}$ to form a link with a contrarian agent $j$ with $P_j > 0$ and minimal $x_j = -x_{max}$ if 
\begin{equation}
\frac{P_i}{P_j} > - k_j
\label{cnrel}
\end{equation}
and we assume that all the neighbours of agents $i$ and $j$ also have the maximal opinion parameters. The striking fact about this relation is that it is always fulfilled in this case, meaning that $i$ would always find formation of links with contrarians acceptable. From the point of view of agent $j$, however, linking to $i$ is only acceptable if 
\begin{equation}
\frac{P_j}{P_i} > \frac{1}{2} \frac{ k_i }{k_j + 1/2},
\label{cnrel2}
\end{equation}
which leads to the very same result as before. i.e. $j$ will not form a link with $i$ unless $P_j > P_i/2$, if $i$ and $j$ are to belong to the same group. Thus the contrarian agents would behave and be treated as normal agents when forming relations, which is surprising considering that their contribution to the total social value of other agents is negative. The latter fact is demonstrated in the simulations with some of the most counterintuitive behaviours of the model, namely, relations between agents being first broken and immediately reinstated. Let us use Eq. (\ref{cutgainij}) to determine, whether the agent $i$ from the previous calculations would benefit from cutting the link with agent $j$, even when expression (\ref{cnrel}) says that $i$ would also form a link with $j$ in the event that such a link did not exist. With the previously stated assumptions, we arrive to the following condition 
\begin{equation}
\frac{P_j}{P_i} > - \frac{ 1 } {k_j - 1}
\label{cncut}
\end{equation}
for the link to be cut. From (\ref{cncut}) we see that unless $k_j = 0$, agent $i$ will cut its ties with the contrarians. Since $k_j \gg 1$ for the largest groups, the inequality (\ref{cncut}) is likely to be true most of the time, leading to links between agents $i$ and $j$ being cut and immediately reformed repeatedly, since inequality~(\ref{cnrel}) also holds. As suggested above, we have observed this behavioural pattern in our simulations, and to some extent it can be observed in Fig. \ref{fig:1bis}, in which it is seen that only the connections between contrarian and normal agents $i$ and $j$ can the signs of $x_i$ and $P_j$ being of opposite signs. It should again be stressed, however, that the calculations above do not take into account the existence of more than one contrarian. Having more contrarians in the system allows them to form links between each other, which has a sizeable effect on the overall structure of the network.

In summary, it could be said that while conditions~(\ref{expla3}) and (\ref{cnrel2}) provide surprisingly well fitting approximations as to how a given agent chooses to link with other agents, the conditions (\ref{cgsimple3}), (\ref{cnrel}) and (\ref{cncut}) (though pointing to the right direction) only give vague qualitative explanations for the behaviour of contrarians and can not be expected to yield precise numerical predictions.

\section{Discussion}

The interpretation of the ranking parameter $a_i$ serves as the key to find possible parallels between our model and the real world. As it describes a single property of an agent, the links between the agents only correspond to exchanges of opinion on whether the agents with larger  $a_i$ are ``better'' than the agents with smaller $a_i$, or vice versa. 

Agents could be considered as being embedded 
a larger social context, and in this context  they could, in principle, have other social connections. In this case, the results presented in the previous Section are best interpreted in terms of echo chambers, which means that  agents prefer such a social hierarchy in which they have better relative rank, and seek to communicate their opinion to others. 
The agents whose ranking parameters are further away from the average, are more vocal in broadcasting their views and gather supporters, since they rank highly in their chosen hierarchy and, therefore, would benefit from their hierarchy becoming more widely accepted. 

On the other, 
the agents with average ranking parameters are much more reluctant to take part in the conversation at all, since they do not rank highly in either of the hierarchies. Then the end result 
for small system sizes is that agents divide themselves according to their ranking pressure into two or more distinct communities supporting opposing hierarchies, in which the agents with similar rankings lump together and refuse to communicate with those that disagree. It is the shutting out of the opposing point of view that makes this system's behaviour reminiscent of echo chambers found in reality. However, with increasing system size the agents develop more nuanced positions on their preferred hierarchies due to mounting social competition, as is seen in the emergence of the contrarians. 

There is, however, an alternative way to interpret the ranking parameter. It could be taken to represent an aggregate of all the social properties of an agent, thereby representing its total standing in the societal status measures. In this case the connections could represent the totality of the agents' social interactions, and the opinion variables the agents' attitude to the (current) state society at large. With this interpretation the rupture between the different communities observed for smaller system sizes would actually represent a real disintegration of the society. This might have implications concerning early human migrations, as they could easily have been influenced by social pressures as well as material needs. If the environmental pressures define a minimum group size necessary for a comfortable life for a tribe, and this minimum is smaller than the limit at which the tribe is forced to be adopting more advanced strategies to enhance social stability, as exemplified by the contrarians in our simulations, the tribe may well split, with splinter groups migrating elsewhere. 

Other than the different economic games, BTH can also shed some light into the well known paradox of value, also known as diamond-water paradox, which refers to the fact that diamonds are far more valued in monetary terms than water, even though water is necessary for life and diamonds not. From the BTH view point the solution to this paradox is obvious: Water, being necessary requirement for life, has to be available in sufficient quantities to all living humans, which means that owning water or its source does not set an individual apart from others, that is, an individual cannot really compare favorably to others on grounds of having water. Diamonds, on the other hand, are relatively rare, and thus cannot be owned by everyone. Therefore, an individual possessing diamonds is compared favourably to others, and so diamonds acquire a relatively high value in comparison to water in the minds of humans, in a very similar manner with which the Veblen goods become valuable. Then BTH, in a sense, contains in itself a natural definition of value, although further work is needed to determine how exactly this status-value relates to other forms of value, such as value derived from usefulness or necessity. 
 
In Section \ref{Model} we only analysed the behaviour of the accepter, since this is straightforward in comparison to predicting the behaviour of the proposer. The experiments on the Ultimatum Game often find that the proposers tend to offer fair shares to accepters, which is easily explained in the context of BTH by the desire of the proposer to have the proposal accepted: the proposers only offer shares that they would accept themselves, and in this way Eq.(\ref{accepted}) also restricts the proposers offers, although it cannot tell the exact amount of money offered. To be able to give a better estimate for the offers one would need to study the learning processes that shape the proposers experience on how uneven treatment people are usually willing to tolerate. 
This is, however, outside the scope of this paper. 

The behaviour of dictators in the dictator game \cite{KKT1986} is somewhat more difficult to analyse using BTH. The dictator game is similar to the Ultimatum game, the only difference being that the other player does not even get to make a choice, and only receives what the first player, or dictator, endows. It has been observed that \cite{kysymys} in this game the dictators tend to be rather generous, which is difficult but not impossible to explain in the context of BTH, if one takes into account the effect of reputation and other ``social goods''. The nature of such influence on the behaviour of the dictator will be studied in a later work. 

However, there are some indications that BTH could very well be applied to the dictator game when all the social effects are taken into account. It has been reported \cite{L2007} that when the rules of the dictator game are modified so that instead of giving money to the other players, the dictator gets to take some or all of the money given to the other players (thus turning the game into a ``taking game''), the dictator's behaviour changes from egalitarian to self serving, i.e. taking often the majority or even all of the available money. From the BTH point of view, the dictator's observed behaviour change can potentially be explained in terms of social norms. In the ordinary dictator game the dictator may still feel bound by the usual norms of the society, while in the ``taking game'' it is encouraged to go against these norms. This sets the ``taker'' apart from the other player in particular, and other members of the society in general.  Hence the dictator feels ``better'' than the others when breaking the norms with impunity, and act on this feeling by taking money from the other players. The fact that BTH can possibly lead to formation of norms as well as rebellion against these norms is well worth of further studies.  

\section{Conclusions}
\label{con}

Relating to the known results of the Ultimatum game, we have formulated a hypothesis explaining the observed behaviour of humans in terms of superiority maximization, or ``better than''-hypothesis, and presented a simple agent-based model to implement this hypothesis. The model describes agents with constant ranking parameters and raises the question whether the agents with larger ranks are ``better'' than agents with smaller ranks or the other way around. 

We have found that the social system produced by our model, features homophily, meaning that agents forming social ties with other agents with similar  ranking parameters, and assortativity, describing the tendency of highly/lowly connected agents forming links with other highly/lowly connected agents. In addition we find community formation, both in terms of there being communities with opposing opinions and in terms of the communities with the same opinion fracturing into smaller ones according to their ranking parameters. Furthermore, we have observed the formation of a hierarchy, in the sense of a connectivity hierarchy being formed on top of the one defined by the ranking parameters, with the agents with extreme ranking parameters presenting higher connectivity than the agents with average ranking parameters.

 Moreover, we have found that the resulting social networks tend to be disconnected for small system sizes, but mostly connected for larger system sizes. This fact may have some relevance for research of early human migrations, hinting of the effects of social pressure in shaping the social network.

\section*{Acknowledgments}

J.E.S. acknowledges financial support from Niilo Helander's foundation, G.I. acknowledges a Visiting Fellowship from the Aalto Science Institute, and K.K. acknowledges financial support by the Academy of Finland Research project (COSDYN) No. 276439 and EU HORIZON 2020 FET Open RIA project (IBSEN) No. 662725. R.A.B. wants to thank Aalto University for kind hospitality during the development of this work. RAB acknowledges financial support from Conacyt through project 799616. We acknowledge the computational resources provided by the Aalto Science-IT project.

\ifx\undefined\BySame
\newcommand{\BySame}{\leavevmode\rule[.5ex]{3em}{.5pt}\ }
\fi
\ifx\undefined\textsc
\newcommand{\textsc}[1]{{\sc #1}}
\newcommand{\emph}[1]{{\em #1\/}}
\let\tmpsmall\small
\renewcommand{\small}{\tmpsmall\sc}
\fi


\begin{thebibliography}{99}

\bibitem{adler}
\textsc{Adler, A.}  (1924) \emph{The Practice and Theory of Individual
  Psychology}. Routledge, Trench and Trubner \& Co, Ltd.

\bibitem{V1899}
\textsc{Veblen, T.}  (1899) \emph{The Theory of the Leisure Class}. Macmillan.

\bibitem{D1949}
\textsc{Duesenberry, J.~S.}  (1949) \emph{Income, savings, and the theory of
  consumer behaviour}. Harvard University Press: Cambridge, Massachusetts.

\bibitem{P1959}
\textsc{Packard, V.}  (1959) \emph{The Status Seekers}. LONGMANS, GREEN AND CO
  LTD6 \& 7 CLIFFORD STREET, LONDON WI.

\bibitem{EGF1999}
\textsc{Eastman, J.~K., Goldsmith, R.~E.  {\small \&} Flynn, L.~R.}  (1999)
  Status Consumption in Consumer Behaviour: Scale Development and Validation.
  \emph{Journal of Marketing Theory and Practice}, \textbf{7}(3), 41–51.

\bibitem{WF1998}
\textsc{Weiss, Y.  {\small \&} Fershtman, C.}  (1998) Social status and
  economic performance: A survey. \emph{European Economic Review}, \textbf{42},
  801--820.

\bibitem{R2008}
\textsc{Rege, M.}  (2008) Why do people care about social status?.
  \emph{Journal of Economic Behavior \& Organization}, \textbf{66}, 233--242.

\bibitem{GS2008}
\textsc{Gaspart, F.  {\small \&} Seki, E.}  (2008) Cooperation, status seeking
  and competitive behaviour: theory and evidence. \emph{Journal of Economic
  Behavior \& Organization}, \textbf{51}, 51--77.

\bibitem{SDT}
\textsc{Sidanius, J.  {\small \&} Pratto, F.}  (1999) \emph{Social Dominance:
  An Intergroup Theory of Social Hierarchy and Oppression}. Cambridge
  University Press.

\bibitem{GSS1982}
\textsc{G\"uth, W., Schmittberger, R.  {\small \&} Schwarze, B.}  (1982) An
  Experimental Analysis of Ultimatum Bargaining. \emph{Journal of Economic
  Behavior \& Organization}, \textbf{3}(4), 367--388.

\bibitem{ES1999}
\textsc{Fehr, E.  {\small \&} Schmidt, K.}  (1999) A theory of fairness,
  competition and cooperation. \emph{Quarterly Journal of Economics},
  \textbf{114}(3), 817--868.

\bibitem{BO2000}
\textsc{Bolton, G.  {\small \&} Ockenfels, A.}  (2000) Erc: A theory of equity,
  reciprocity and competition. \emph{The American Economic Review},
  \textbf{90}, 166--193.

\bibitem{L2005}
\textsc{Luttmer, E. F.~P.}  (2005) Neighbors as Negatives: Relative Earnings
  and Well Being. \emph{Quarterly Journal of Economics}, pp. 963--1002.

\bibitem{ISS2008}
\textsc{Izuma, K., Saito, D.~N.  {\small \&} Sadato, N.}  (2008) Processing of
  Social and Monetary Rewards in the Human Striatum. \emph{Neuron},
  \textbf{58}(2), 284--294.

\bibitem{ZTCB2008}
\textsc{Zink, C.~F., Tong, Y., Chen, Q., Bassett, D.~S., Stein, J.~L.  {\small
  \&} Meyer-Lindenberg, A.}  (2008) Know Your Place: Neural Processing of
  Social Hierarchy in Humans. \emph{Neuron}, \textbf{58}(2), 273--283.

\bibitem{A1981}
\textsc{Axelrod, R.  {\small \&} Hamilton, W.~D.}  (1981) The Evolution of
  Cooperation. \emph{Science}, \textbf{211}, 1390--1396.

\bibitem{A1986}
\textsc{Axelrod, R.}  (1981) An Evolutionary Approach to Norms. \emph{American
  Political Science Review}, \textbf{80}(4), 1095--1111.

\bibitem{Y1993}
\textsc{Young, H.~P.}  (1993) The evolution of conventions.
  \emph{Econometrica}, \textbf{61}, 57--84.

\bibitem{Y1995}
\textsc{\BySame{}}  (1995) The economics of convention. \emph{The Journal of
  Economic Perspectives}, \textbf{10}, 105--122.

\bibitem{CWM2005}
\textsc{Centola, D., Willer, R.  {\small \&} Macy, M.}  (2005) The Emperor's
  Dilemma: A Computational Model of Self-Enforcing Norms. \emph{American
  Journal of Sociology}, \textbf{10}(4), 1009--1040.

\bibitem{FOCN2016}
\textsc{Fagundes, M.~S., Ossowski, S., Cerquides, J.  {\small \&} Noriega, P.}
  (2016) Design and evaluation of norm-aware agents based on Normative Markov
  Decision Processes. \emph{International Journal of Approximate Reasoning},
  \textbf{78}, 33--61.

\bibitem{T1988}
\textsc{Thaler, R.~H.}  (1988) Anomalies: The Ultimatum Game. \emph{The Journal
  of Economic Perspectives}, \textbf{2}(4), 195--206.

\bibitem{GT1990}
\textsc{G\"uth, W.  {\small \&} Tietze, R.}  (1990) Ultimatum bargaining
  behavior: A survey and comparison of experimental results. \emph{Jornal of
  Econonomic Psychology}, \textbf{11}(3), 417--449.

\bibitem{kysymys}
\textsc{J., H., Boyd, R., Bowles, S., Camerer, C., Fehr, E.  {\small \&}
  Gintis, H.}  (2004) \emph{Foundations of Human Sociality: Economic
  Experiments and Ethnographic Evidence from Fifteen Small-Scale Societies}.
  Oxford University Press.

\bibitem{metalyysi}
\textsc{Oosterbeek, H., Sloof, R.  {\small \&} van~de Kuilen, G.}  (2004)
  Cultural Differences in Ultimatum Game Experiments: Evidence from a
  Meta-Analysis. \emph{Experimental Economics}, \textbf{7}(2), 171--188.

\bibitem{FHSS1994}
\textsc{Forsythe, R., Horowitz, J., Savin, N.~E.  {\small \&} Sefton, M.}
  (1994) Fairness in Simple Bargaining Experiments. \emph{Games and Economic
  Behavior}, \textbf{6}, 347--369.

\bibitem{BZ1995}
\textsc{Bolton, G.~E.  {\small \&} Zwick, R.}  (1995) Anonymity versus
  punishments in ultimatum bargaining. \emph{Games and Economic Behavior},
  \textbf{10}(1), 95--121.

\bibitem{IKKB2009}
\textsc{I{\~n}iguez, G., Kert{\'e}sz, J., Kaski, K.~K.  {\small \&} \&~Barrio,
  R.~A.}  (2009) Opinion and community formation in coevolving networks.
  \emph{Physical Review E}, \textbf{80}, 066119.

\bibitem{PPZ}
\textsc{Piraveenan, M., Prokopenko, M.  {\small \&} Zomaya, A.~Y.}  (2008)
  Local assortativeness in scale-free networks. \emph{Europhysics Letters},
  \textbf{84}(2), 28002.

\bibitem{KKT1986}
\textsc{Kahneman, D., Knetsch, J.~L.  {\small \&} Thaler, R.~H.}  (1986)
  Fairness and the Assumptions of Economics. \emph{The Journal of Business},
  \textbf{59}(4), 285--300.


\bibitem{L2007}
\textsc{List, J.~A.}  (2007) On the Interpretation of Giving in Dictator Games.
  \emph{Journal of Political Economy}, \textbf{115}(3), 482--493.

\end{thebibliography}

\end{document}